# Simultaneous ρ-v inversion by extended Gel'fand-Levitan-Marchenko solution in layered acoustic media for oblique incidence

*Ru-Shan Wu[1†] and Huijing He[1]*

[1] *Modeling and Imaging Laboratory, Earth and Planetary Sciences, University of California, Santa Cruz, Santa Cruz, California 95064, USA*
[†] *Corresponding author: rwu@ucsc.edu*

**Abstract**

Original GLM (Gel'fand-Levitan-Marchenko) theory is for scattering potential recovery of Schrödinger equation. In this paper, we formulate the GLM impedance solution of oblique incidence for simultaneous inversion of velocity and density including boundary jumps based on acoustic wave equation. In addition to the use of reflection amplitude, traveltime information of reflection arrivals is also utilized for inversion based on the focusing principle. Numerical tests demonstrate the recovery of velocity and density, verified the validity of the theory and method. The theory of Schrödinger impedance inverse-scattering studied in this paper is closely related to the direct envelope inversion.

## 1. Introduction

GLM (Gel'fand-Levitan-Marchenko) theory [1,2] was introduced originally for scattering potential recovery of one-dimensional time-independent Schrödinger equation. Later it has been applied to the inverse-scattering for potential and impedance [3]. Ware and Aki [4] applied the theory to solve the 1-D elastic equation (for horizontal shear wave), but the solution is limited to the case of smooth impedance variation. Berryman and Greene [5] introduced an integral for the K-operator to recover the low-wavenumber component of impedance. Coen [6] formulated the GLM equation for oblique incidence to recover both velocity and density. The reduced GLM equation becomes an equation for refractive index, and the solution is also limited to smooth media. Howard [7] developed a vector Marchenko equation based on the first order differential equation of motion and also gave a solution of extracting density and velocity from reflection amplitudes of two incident angles. Wu [8] applied the GLM theory and method to 1-D acoustic medium inversion including sharp boundaries by using the symbolic operation based on distribution theory and connect the method to the direct envelope inversion (DEI). In this paper, we derive the multi-dimensional Schrödinger impedance equation for layered acoustic media including oblique incidence for ρ-v inversion. Coen [6] first derived the equation for refractive index through a density-variable transform, then change the refraction-index equation into a Schrödinger equation by Liouville transform. Here we applied the Liouville transform first using the effective velocity for oblique incidence and then transform the acoustic wave equation into a Schrödinger impedance equation for acoustic media including sharp discontinuities. Different from Coen [6] and Howard [7], where only reflection amplitude is used for inversion, we use both reflection amplitude and traveltime of reflection arrivals for impedance and velocity determination, which has a potential to stabilize the inversion. We conduct numerical tests for a layered medium containing an impedance layer with both velocity and density perturbations and a pure density perturbation layer. The results show that the method can recover both the velocity and density of layered media from surface reflection measurements.

## 2 Schrödinger impedance equation in layered acoustic media for oblique incidence

In a general three-dimensional acoustic media, the wave equation can be written as

$$\rho(\mathbf{x})\nabla \cdot \left(\frac{1}{\rho(\mathbf{x})}\nabla p(\mathbf{x},t)\right) - \frac{1}{c^2(\mathbf{x})}\frac{\partial^2}{\partial t^2}p(\mathbf{x},t) = 0. \quad (1)$$

In layered acoustic media, the density $\rho(z)$ and velocity $c(z)$ are only z-dependent, but the pressure wavefield $p(\mathbf{x},t)$ is a function of time and three-dimensional space. In this case the first term in above equation can be reduced to

$$\rho(\mathbf{x})\nabla \cdot \left(\frac{1}{\rho(\mathbf{x})}\nabla p(\mathbf{x},t)\right) = \nabla^2 p(\mathbf{x},t) - \frac{1}{\rho(\mathbf{x})}\nabla \rho(\mathbf{x}) \cdot \nabla p(\mathbf{x},t)$$
$$= \nabla^2 p(\mathbf{x},t) - \frac{1}{\rho(z)}\frac{\partial}{\partial z}\rho(z)\frac{\partial}{\partial z}p(\mathbf{x},t). \quad (2)$$

Therefore, acoustic equation for 1-D media with 3-D wavefield becomes

$$\nabla^2 p(\mathbf{x},t) - \frac{1}{\rho(z)}\frac{\partial}{\partial z}\rho(z)\frac{\partial}{\partial z}p(\mathbf{x},t) - \frac{1}{c^2(z)}\frac{\partial^2}{\partial t^2}p(\mathbf{x},t) = 0. \quad (3)$$

By Fourier transform, the above equation can be written in frequency-domain,

$$\nabla^2 p(\mathbf{x},\omega) - \frac{1}{\rho(z)}\frac{\partial}{\partial z}\rho(z)\frac{\partial}{\partial z}p(\mathbf{x},\omega) + \frac{\omega^2}{c^2(z)}p(\mathbf{x},t) = 0. \quad (4)$$

For a plane wave with incident angle $\theta_0$

$$p_0(\mathbf{x},\omega) = \exp[ik_0(x\sin\theta_0 + z\cos\theta_0)], \quad (5)$$

where $k_0 = \omega/c_0$ with $c_0$ as the velocity near the surface, the general solution in the upper-half space (free space) is

$$p(\mathbf{x},\omega) = \exp(ik_0 x\sin\theta_0)$$
$$\times[\exp(ik_0 z\cos\theta_0) + R(\omega,\theta_0)\exp(-ik_0 z\cos\theta_0)], \quad z < 0 \quad (6)$$

where $R(\omega,\theta_0)$ is the reflection coefficient for oblique incidence. We know that the horizontal component of the Laplacian $\nabla_x^2 = \frac{\partial^2}{\partial x^2} = -k_0^2 \sin^2\theta_0$ will keep constant, since the horizontal slowness $\sin\theta_0/c_0$ is an invariant in 1-D media (Snell's law). Substitute the above equation into Eq. (4) resulting in

$$\left[\left(\frac{\partial^2}{\partial z^2} - \frac{\omega^2}{c_0^2}\sin^2\theta_0 + \frac{\omega^2}{c^2(z)}\right) - \frac{1}{\rho(z)}\frac{\partial}{\partial z}\rho(z)\frac{\partial}{\partial z}\right]p(\mathbf{x},\omega) = 0. \tag{7}$$

Above equation can be written into a 1-D acoustic wave equation

$$\left[\left(\frac{\partial^2}{\partial z^2} + \frac{\omega^2}{c_\theta^2(z)}\right) - \frac{1}{\rho(z)}\frac{\partial}{\partial z}\rho(z)\frac{\partial}{\partial z}\right]p(z,\omega) = 0. \tag{8}$$

with an apparent velocity $c_\theta$ depending on the incident angle

$$c_\theta(z) = \frac{c(z)}{\sqrt{1-[c(z)/c_0]^2 \sin^2\theta_0}}. \tag{9}$$

In the case of pre-critical reflection, $c_\theta$ has always limited value. For normal incidence, $\theta_0 = 0$ and $c_\theta(z) = c(z)$. In the time-domain, Eq. (7) becomes

$$\left[\left(\frac{\partial^2}{\partial z^2} - \frac{1}{c_\theta^2(z)}\frac{\partial^2}{\partial t^2}\right) - \frac{1}{\rho(z)}\frac{\partial}{\partial z}\rho(z)\frac{\partial}{\partial z}\right]p(z,\omega) = 0. \tag{10}$$

It has been shown that 1-D acoustic wave equation such as Eq. (10) can be transformed into a Schrödinger equation [4,8],

$$\left(\frac{\partial^2}{\partial \varsigma_\theta^2} - \frac{\partial^2}{\partial t^2}\right)\psi(\varsigma_\theta,t) = q_\theta(\varsigma_\theta)\psi(\varsigma_\theta,t) + s(\varsigma_\theta,t),$$

$$q_\theta(\varsigma_\theta) = \left(\frac{1}{\eta_\theta(\varsigma_\theta)}\frac{\partial^2}{\partial \varsigma_\theta^2}\eta_\theta(\varsigma_\theta)\right), \tag{11}$$

where $q_\theta$ is the scattering potential and $s(\varsigma_\theta,t)$ is the real source (primary source). To derive the above Schrödinger equation, both independent and dependent variable transforms have been applied. The independent variable $z$ has been changed to a travel-time related variable $\varsigma_\theta$ by the Liouville transform,

$$d\varsigma_\theta = dz / c_\theta(z) \tag{12}$$

and the dependent variable (pressure field $p$) was changed to an energy-flux normalized field $\psi$ through

$$\psi(\varsigma_\theta,t) = p(\varsigma_\theta,t)\eta_\theta(\varsigma_\theta), \tag{13}$$

where $\eta_\theta(\varsigma_\theta) = 1/\sqrt{\rho(\varsigma_\theta)c_\theta(\varsigma_\theta)}$ is the impedance function with $\rho c$ defined as the impedance. Note that the equivalent potential in the impedance Schrödinger equation has a second order derivative in the operation. Therefore, the original impedance inversion applying the GLM theory has more severe limitation than the potential inversion, and the impedance function can be inverted was limited to a smooth one. Wu [8] has introduced symbolic operation based on the distribution theory, and then applied to the GLM equation for impedance inversion with discontinuities. It was also pointed out by Berryman and Greene [5] that the impedance equation in Eq. (11) observes the zero-frequency Schrödinger equation, and therefore can be solved by the zero-frequency Jost solution of Schrödinger equation, leading to a impedance solution composed by an integral of the GLM kernel function $K(\varsigma_\theta,t)$,

$$\eta_\theta(\varsigma_\theta) = \eta_\theta(0)\left[1 + \int_{-\varsigma_\theta}^{\varsigma_\theta} K(\varsigma_\theta, t)dt\right]. \tag{14}$$

The kernel $K(\varsigma_\theta, t)$ can be obtained by solving the GLM integral equation [1,2],

$$K(\varsigma_\theta, t) = -R_\theta(t + \varsigma_\theta) - \int_{-\varsigma_\theta}^{\varsigma_\theta} K(\varsigma_\theta, \tau)R_\theta(t + \tau)d\tau, \tag{15}$$

$$\varsigma_\theta \geq 0, \ -\varsigma_\theta \leq t \leq \varsigma_\theta$$

where $R_\theta(t)$ is the reflection record (reflection time series) on the surface for oblique incidence. For details of GLM theory, see Berryman and Greene [5] and Wu [8].

## 3 Density and velocity inversion after obtaining the GLM solutions for oblique incidence

After obtain impedance functions $\eta_\theta(\varsigma_\theta)$ at least for two angles, one for normal incidence $\eta(\varsigma)$, one for incident angle $\theta$, there will be enough information to invert both density and velocity by GLM inverse-scattering theory. Coen [6] formulated the GLM equation for refraction index, and solve the density first by subtracting the results for two angles. Here we formulate the GLM equation for impedance and then solve the velocity distribution first. In this way, we hope the physical meaning will be more transparent and the velocity inversion can be more stable.

If we measure reflection responses for both normal incidence R(t) and oblique incidence $R_\theta(t)$, then we can obtain the corresponding impedance function $\eta(\varsigma)$ and $\eta_\theta(\varsigma_\theta)$. Based on the two recovered impedance curves, we can use both the amplitude ratio and the traveltime ratio of angle-dependent reflections for velocity inversion.

1) **Impedance ratio method of velocity estimation**. Since the near-surface velocity $c_0$ is assumed known, so the first layer's position $z_1$ can be calculated. Then velocity of the first layer can be obtained by taking the ratio of the two impedance functions:

$$\frac{\eta_\theta^2(z_1^+)}{\eta^2(z_1^+)} = \frac{\rho(z_1^+)c(z_1^+)}{\rho(z_1^+)c_\theta(z_1^+)} = \frac{c(z_1^+)}{c_\theta(z_1^+)} = \sqrt{1 - \frac{c^2(z_1^+)\sin^2\theta_0}{c_0^2}}. \tag{16}$$

Since we adopt the discontinuous layer model, so the impedance or velocity structures are involved with step functions, $z_1^+$ in above equation denotes the down-side (or the "right-hand") side of the step function. In this paper, when we specify the velocity or other parameter with a position variable, it always means the next layer starting from the given position. So we will use $z_n$ instead of $z_n^+$ for nth layer position. From Eq. (16), we see that the $\rho-$ dependence is removed by taking the impedance ratio of two incident angles. Physically we understand that angle-dependence of refraction is a forward-scattering phenomenon and depends only on velocity structure. Then the velocity of the first layer is obtained as

$$c(z_1) = \frac{c_0}{\sin\theta_0}\sqrt{1-\frac{\eta_\theta^4(z_1)}{\eta^4(z_1)}} = \frac{c_0}{\sin\theta_0}\sqrt{1-\frac{\eta_\theta^4(z_1(\varsigma_\theta))}{\eta^4(z_1(\varsigma))}}, \quad (17)$$

where $z_1(\varsigma)$ are calculated by the inverse Liouville transform,

$$z_1(\varsigma) = z_0 + \int_{\varsigma_0}^{\varsigma} c(\varsigma')d\varsigma'. \quad (18)$$

For each layer, the density $\rho(z)$ is easy to obtain from the impedance function with known velocity by

$$\rho(z) = \frac{1}{c(z)\eta^2(z)}. \quad (19)$$

After inverting the velocity of the first layer, it's layer thickness can be calculated by the inverse Liouville transform as.

$$\Delta z_1 = c(z_1)\Delta\varsigma_1 = c_1\Delta\varsigma_1. \quad (20)$$

For the nth layer, the velocity and thickness can be obtained as

$$c_n = \frac{c_0}{\sin\theta_0}\sqrt{1-\frac{\eta_\theta^4(z_n)}{\eta^4(z_n)}}, \quad (21)$$

$$\Delta z_n = c_n\Delta\varsigma_n.$$

The above procedure is only based on the reflection amplitude-ratio and may suffer from the error propagation for layer position estimation. In order to avoid the error propagation problem, we can calculate the layer velocity using the traveltime data of the angle-dependent reflection arrivals which are independent from the amplitude data.

**2) Traveltime method for velocity estimation based on focusing principle.** Traveltimes between layer *n* and layer *n+1* can be measured from the impedance curves as solutions of GLM equation for both normal and oblique incidences,

$$\Delta\varsigma(n) = \varsigma(n+1) - \varsigma(n), \\ \Delta\varsigma_\theta(n) = \varsigma_\theta(n+1) - \varsigma_\theta(n). \quad (22)$$

The corresponding spatial distances are

$$\Delta z(n) = c_n\Delta\varsigma(n), \\ \Delta z_\theta(n) = c_n\Delta\varsigma_\theta(n)/\cos\theta_n. \quad (23)$$

By the focusing principle, the pulses from normal incidence and oblique incidence should focus to the same spatial location, so $\Delta z(n) = \Delta z_\theta(n)$ should hold, resulting in

$$\frac{\Delta\varsigma_\theta(n)}{\Delta\varsigma(n)} = \cos\theta_n = \sqrt{1-\frac{c_n^2}{c_0^2}\sin^2\theta}. \quad (24)$$

Then the velocity of the nth layer can be determined from the travel time ratio as

$$\bar{c}_n = \frac{c_0}{\sin\theta_0}\sqrt{1-\left(\frac{\Delta\varsigma_\theta(n)}{\Delta\varsigma(n)}\right)^2}, \quad (25)$$

where $\bar{c}_n$ is the nth layer velocity estimated from the traveltime data. Compare to the velocity estimate from the amplitude data (21), this $\bar{c}_n$ estimate may have different noise-resistance property and do not suffer from the error propagation problem. In practical application, data from more incident angles can be used to further stabilize the inversion and reduce inversion errors. An optimization scheme may be set up for the inversion using redundant data.

## 4 Numerical tests

To demonstrate the validity of the proposed method, we consider a layered half space with an impedance layer having both velocity and density perturbations and a pure density perturbation layer, as shown in Fig. 1. Note that in the example, the bottom half-space has different parameters from the near-surface layer.

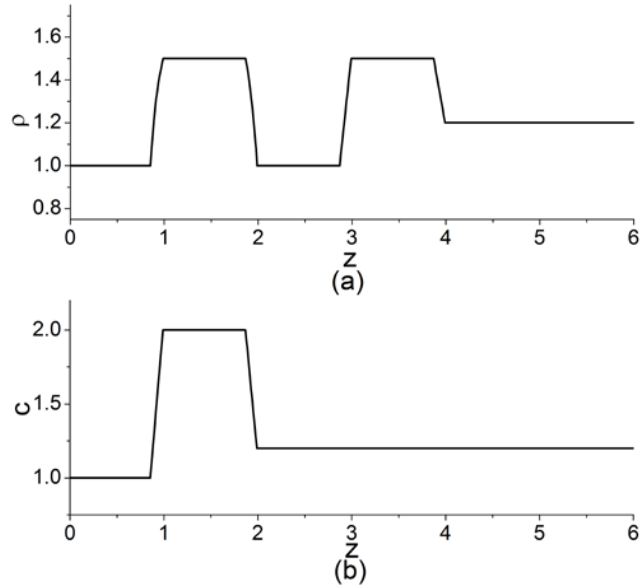

FIG. 1. Original model: (a) density, (b) velocity

The reflection spectrum of the inhomogeneous medium is calculated by first subdividing the structure into a series of thin layers and then using the standard transfer matrix method. The frequency band for generating the synthetic seismograms in our calculation is from 0 to 41 Hz. The time-domain impulse reflection series for normal incidence R(t) and for oblique incidence $R_\theta(t)$ are then generated by the fast Fourier transform (FFT) algorithm. This completes the forward simulation of the scattering problem. The normal and oblique reflection signals are shown in Fig. 2.

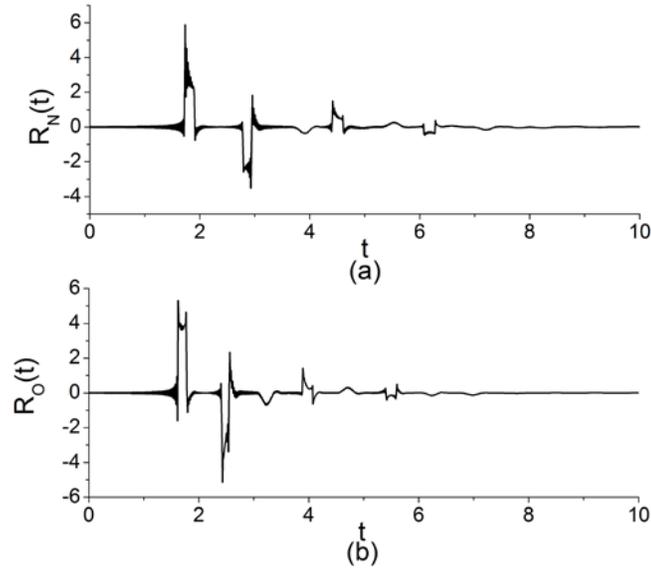

FIG. 2. Time domain reflection signals: (a) normal incidence, (b) oblique incidence

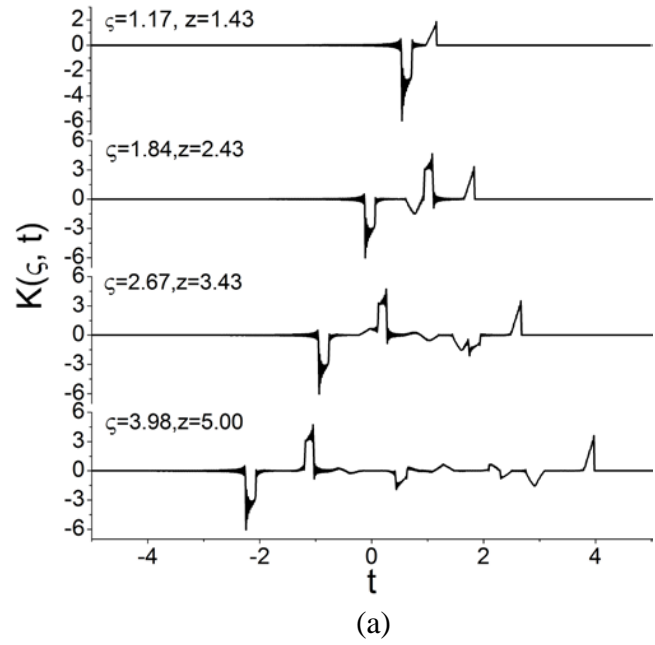

(a)

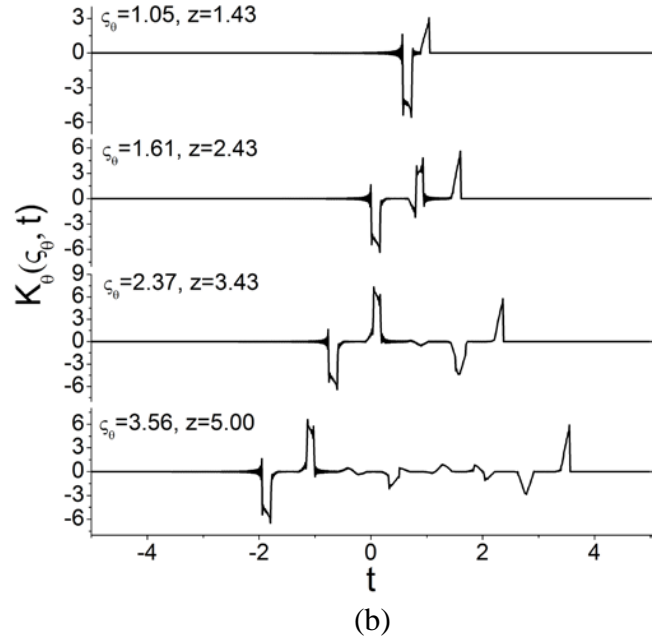

(b)

FIG. 3. Solutions of the GLM equation: (a) K($\varsigma$, t), (b) K$_\theta$($\varsigma_\theta$, t)

The kernel functions of the normal and oblique incidence, $K(\varsigma,t)$ and $K_\theta(\varsigma_\theta,t)$ are obtained by solving the corresponding GLM equation. An iterative technique proposed by Ge [9] is used to solve the integral equations. For the treatment of the boundary jumps using generalized function theory, see Wu [8,10]. The iteration uses the reflection time series as the initial guess, and after 50 times of iteration, it converges to the correct results. Fig. 3 shows the cross sections of the kernels $K(\varsigma,t)$ and $K_\theta(\varsigma_\theta,t)$ at four different depths: z=1.43, 2.43, 3.43, and 5. The calculated kernel function shown as a time series at each cross section corresponds to the focusing wavefield which focuses at the given depth, as explained by Rose [11].

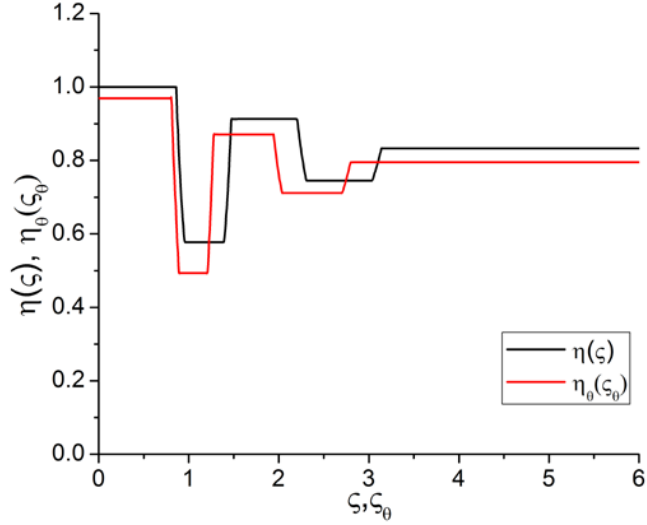

FIG. 4. Reconstructed impedance functions: η(ς) (black line) and η_θ(ς_θ) (red line)

$K(\varsigma,t)$ and $K_\theta(\varsigma_\theta,t)$ are then used to calculate the acoustic impedance functions $\eta(\varsigma)$ and $\eta_\theta(\varsigma_\theta)$ following the recipe given by Berryman and Greene [5], see Eq. (14). The calculated impedance functions are shown in Fig. 4.

Finally, the density and velocity profiles are reconstructed according to the method described in Sec. 3. It is worth mentioning that the depth increments of the normal and oblique incidences are different, see Eq. (12). Thus, to facilitate the property inversion, here we use the linear interpolation of the nearest neighboring points to obtain the acoustic impedance functions at the same spatial coordinate. Refined grids are necessary to improve the accuracy of the reconstruction procedure. For sharp boundaries we need to pay attention to singularities. Here we use the boundary layer integral to remove the influence of singularities. The reconstructed density and velocity profiles are shown in Fig. 5. For the oblique incidence, we considered two incident angles, one is 20° and the other is 15°. It is seen that for both cases the density and velocity profiles are recovered with satisfactory accuracy. At the interfaces between different layers there are still small artifacts. Exploring new strategies to remove these artifacts and study the stability problem are among the research subjects in our future work.

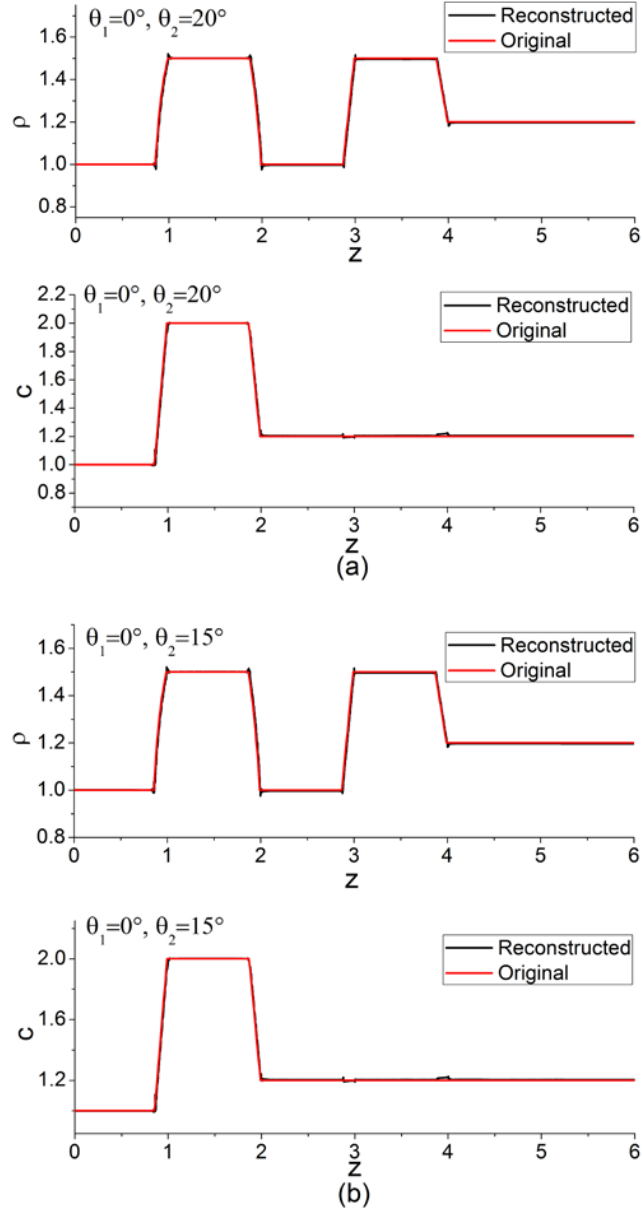

FIG. 5. Reconstructed results using amplitude ratio: (a) results for incident angle $0°$ and $20°$, (b) results for $0°$ and $15°$.

As discussed in Sec. 3, the medium parameters can also be reconstructed by using the "traveltime" information in the obtained impedance functions $\eta(\varsigma)$ and $\eta_\theta(\varsigma_\theta)$. The discontinuities in the impedance functions are displayed in the form of step functions. In order to locate the interfaces between different layers and thus, calculate the traveltime more accurately, we further calculate and plot the derivatives of the impedance functions, see Fig. 6. The spikes in the figure represent the transition layers, and the traveltime of each interface can be fixed by the location of the corresponding inflection point in the spikes. With the traveltime information, the velocity of each major layer is reconstructed by using Eq. (25). The thickness of each layer is further obtained via the inverse Liouville transform. We need to mention that the thickness of the transition layers is calculated by taking the product of the traveltime of the transition layer and its effective velocity, where the latter is obtained by replacing the traveltime in Eq. (25) with the width of the corresponding spikes. Once the velocity is available, the density is reconstructed by using Eq. (19). The reconstruction results for incident angles of 0° and 20° are shown in Fig. 7. It is seen that both the velocity and density are recovered with satisfactory accuracy.

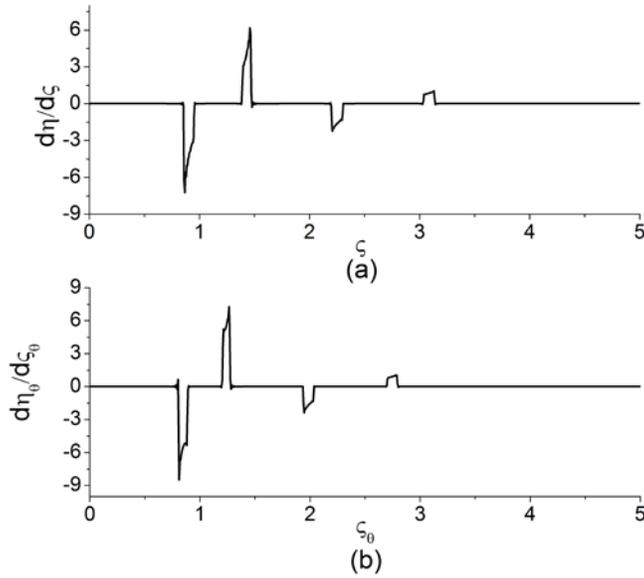

FIG. 6. Derivatives of the acoustic impedance functions:
(a) Normal incident; (b) Oblique incident (20°)

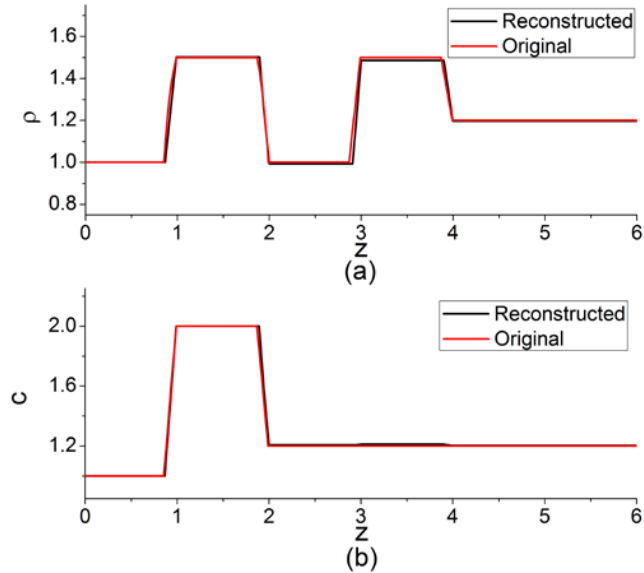

FIG. 7. Reconstructed results using traveltime information:
(a) density, (b) velocity

## 5 Conclusion

We propose a simultaneous rho-v inversion method for layered acoustic media based on GLM (Gel'fand-Levitan-Marchenko) theory for oblique incidence. We derived the GLM impedance equation in acoustic layered media for oblique incidence and introduced two methods of using the amplitudes and arrivals times of reflection arrivals for velocity and density simultaneous inversion. Numerical tests demonstrate the feasibility of the theory and method. Noise resistance and inversion stability will be the next topic of study.

## Acknowledgments

This work is supported by WTOPI (Wavelet Transform On Propagation and Imaging/Inversion for seismic exploration) Research Consortium and other funding resources at the Modeling and Imaging Laboratory, University of California, Santa Cruz.